\pdfoutput=1

 \documentclass[final,3p,times,twocolumn]{elsarticle}

\usepackage{amssymb}
\usepackage{bm}
\usepackage{amsmath}
\usepackage{widetext}
\usepackage{xcolor}

\journal{Nuclear Instruments and Methods in Physics Research A}

\newcommand\blfootnote[1]{%
  \begingroup
  \renewcommand\thefootnote{}\footnote{#1}%
  \addtocounter{footnote}{-1}%
  \endgroup
}

\begin{document}

\begin{frontmatter}



\title{A method to measure the integral vertical intensity and angular distribution of atmospheric muons with a stationary plastic scintillator bar detector}


\author{Bryan Olmos Y\'a\~nez \corref{cor1} }
\ead{olmos.bryan@ciencias.unam.mx}
\author{Alexis A. Aguilar-Arevalo}
\ead{alexis@nucleares.unam.mx}

\cortext[cor1]{Corresponding author}

\address{Instituto de Ciencias Nucleares. Universidad Nacional Aut\'onoma de M\'exico \\
 Circuito Exterior, Ciudad Universitaria, Apartado postal 70-543, 04510  CDMX, M\'exico}

\begin{abstract}
\noindent We present a method to measure the integral vertical intensity  and angular distribution of atmospheric muons using a stationary thick plastic scintillator bar detector with PMTs at both of its ends over exposures of a few hours in a laboratory setting.
The total flux is obtained from the event rate and the angular distribution is inferred from the distribution of the recorded pulse charges, which are correlated with the track length inside the plastic scintillator.
A muon generator algorithm was developed and used together with a simple simulation including the effects of the detector resolution, nonlinear saturation of the PMTs, and laboratory building coverage.
As a proof of principle we made a measurement assuming a standard $\cos^n\theta$ omnienergetic angular distribution with azimuthal isotropy and different models for the energy spectrum of the source. 
Using measurements at 5 different azimuthal orientations, we measure an average exponent $n=(1.90\pm 0.06({\rm stat})\pm 0.10({\rm syst}))$, and an average vertical intensity of $I_0=(101.2\pm 1.8({\rm stat})\pm 5.5({\rm syst}))$ $~{\rm m}^{-2}{\rm s}^{-1}{\rm sr}^{-1}$ at the location with geographic coordinates 19.33$^\circ$N~99.19$^\circ$W, altitude 2,268~m above sea level, and geomagnetic cut-off of 8.2~GV.
\end{abstract}




\end{frontmatter}


\section{\label{sec:intro}Introduction}

\blfootnote{\textcopyright~2020. This manuscript version is made available under the CC-BY-NC-ND 4.0 license http://creativecommons.org/licenses/by-nc-nd/4.0/}
\hspace{-0.2cm}
Atmospheric muons are produced in the decay of secondary particles (mostly pions an kaons) created in the interaction of primary cosmic rays, such as protons and heavier nuclei, with the Earth's atmosphere at high  altitudes, typically of the order of $15$~km. At sea level muons ($\mu^+$, $\mu^-$) are the most abundant charged particles in the extended air showers created by cosmic rays, and reach the ground with a mean energy of approximately $4$~GeV. At various latitudes and altitudes it has been observed that the muon angular distribution is well described by  $I(\theta) = I_0 \;\cos^n\theta$, where $I_0$ is the integrated vertical flux intensity (m$^{-2}$s$^{-1}$sr$^{-1}$), and $\theta$ is the zenith angle of the incident muon. Both, $I_0$ and the exponent $n$ depend on many factors, such as altitude, latitude, geomagnetic cut-off, and, over shorter timescales, solar activity and atmospheric conditions. A summary of measurements can be found in \cite{pethuraj:2017}.
 For muons with energies above 1~GeV at typical sea level altitude locations, the integral vertical intensity is about 70~${\rm m}^{-2}{\rm s}^{-1}{\rm sr}^{-1}$ \cite{depascale:1993,grieder:2001}, although some measurements suggest a value that is lower by 10-15\% \cite{kremer:1999, haino:2004, archard:2004}. Significantly higher values have been reported \cite{fauth:2007} within the South Atlantic Anomaly \cite{augusto:2010} where the Earth's magnetic field is strongly reduced. In the flat-atmosphere approximation, the exponent $n$, can be shown to be one unit lower than that of the assumed primary cosmic ray spectrum ($\sim$$E^{-(n+1)}$) \cite{shukla:2018}, and is expected to lie around the value of 2 for energies $>1$ ~GeV.

Reliable data on the energy and angular distribution of atmospheric muons at a variety of geographic locations exist for $E>200$~MeV \cite{boezio:2003,liu:2003}. These data, together with a large body of measurements of other components of the cosmic radiation reaching Earth, have been used to validate simulations of the production of atmospheric showers which can be used to predict the muon flux at virtually any location on the Earth's surface \cite{corsika:1998.book,expacs1:2016,expacs2:2015, quarm:2004, quarm:2006}.

Cosmogenic muons have been either used or proposed for many applications, for example: to search for hidden chambers in the Egyptian pyramids \cite{alvarez:1970}, in the non-destructive assessment of radioactive  materials stored in sealed containers \cite{chatzidakis:2017}, to determine the location of high density material in reactor meltdowns like Chernobyl and Fukushima \cite{perry:2003,miyadera:2013}, to image the interior of volcanoes and study the evolution of their internal structure \cite{tioukov:2019}. Muon tomography (muography) offers great potential as a new technology for the evaluation and screening of materials and structures beyond the conventional x-ray, and gamma techniques \cite{morris:2008}.


In this work, we implement a simple method to extract the vertical muon intensity and angular distribution (exponent $n$) for positive plus negative muons ($\mu^+ + \mu^-$) at a given location on the Earth's surface using a stationary, thick scintillator bar detector coupled to photomultiplier tubes (PMTs) at both of its ends. The thickness of the scintillator provides access to sampling the angular distribution of the incoming muon flux, which constrains its absolute normalization, and generic assumptions about the detector response allow for a good description of the spectrum shape without the need of a detailed simulation of the scintillation processes, requiring relatively little computing power.

The method relies on a simple muon generator algorithm suitable for simulating the cosmogenic muon flux raining over a detector of finite size situated on the Earth surface, and a basic Geant4 simulation of the energy deposited by muons on the scintillator material, as well as the coverage provided by the building where the measurement is performed. 

The paper is organized as follows. 
In Section \ref{sec:smith}, we briefly describe the phenomenological model of the muon flux reaching the ground to be used as a benchmark in our study.
In Section \ref{sec:generator}, we describe an algorithm to generate atmospheric muons reaching a plastic scintillator bar on the surface of the Earth.
In Section \ref{sec:simulation}, details of the simulation of the detector are given, and simple validation tests are performed.
In Section \ref{sec:exper}, we go over  the experimental setup used to acquire the energy spectra of atmospheric muons reaching the ground with a plastic scintillator bar.
In Section \ref{sec:fits}, we fit the model of the simulation to the experimental spectrum including energy resolution and non-linearity effects.
In Section \ref{sec:muon-int}, we extract the muon vertical intensity and angular distribution, and explore variations due to systematic uncertainties. 
In Section \ref{sec:closure-tests}, we perform two closure tests to validate the robustness of the fitting procedure.
Finally, in Section \ref{sec:var-orient}, we show the results of applying the method to data taken with various azimuthal orientations of the scintillator bar.
The conclusions are discussed in Section \ref{sec:conclusions}.

\section{\label{sec:smith} Muon flux model}

We will use the phenomenological model of Smith-Duller \cite{smith-duller:1959} to describe the differential intensity of atmospheric muons as a benchmark. 
This simple model captures the main characteristics of experimentally measured energy and angular distributions of cosmogenic muons on the ground, and has previously been used for similar purposes \cite{chatzidakis:2015}.
The model calculates the number of muons ($N$) per unit time ($t$), area ($A$), energy ($E_\mu$), and solid angle ($\Omega$), reaching the ground at atmospheric depth $y_0$ and air density $\rho_0$, originating from the decay of pions (rest mass and lifetime $m_\pi$ and $\tau_0$, respectively) produced at high altitudes in the Earth's atmosphere as follows

\begin{equation}
\frac{dN}{dA\;d\Omega\;dt\;dE_\mu} (E_\mu, \theta)= 
\frac{A \; E_\pi^{-k} P_\mu \; \lambda_\pi \; b \; j_\pi}{E_\pi \cos\theta + b \;j_\pi},
\label{eq:sd:Nmu}
\end{equation}

\noindent
where $A$ is a normalization parameter; $k=8/3$ is the spectral index inherited from pion production; $E_\pi$ is the parent pion energy prior to its decay, $\lambda_{\pi}=120~{\rm g/cm}^2$ is the pion absorption mean free path at high energies, $j_{\pi}=y_0 m_{\pi}c/\tau_0 \rho_0$, $b=0.771$ is a numerical factor introduced to correct the isothermal atmosphere approximation at high altitudes, and $P_\mu$ is the probability that a muon coming down at zenith angle $\theta$ reaches the Earth's surface without decaying, approximated by

\begin{equation}
P_\mu = 
\left[ 
0.100 \cos\theta\left( 1- \frac{a(y_0\sec\theta-100)}{r E_\pi} \right)
\right]^\frac{B_\mu}{(rE_\pi+100 a)\cos\theta} ,
\label{eq:sd:Pmu}
\end{equation}

\noindent
where $r=0.76$ is the fraction of the parent pion energy taken by the muon (assumed constant), $B_\mu = b_\mu y_0 m_\mu c / \tau_{\mu 0} \rho_0$, with $m_\mu$ and $\tau_{\mu 0}$ the muon mass and lifetime at rest, respectively, $b_\mu=0.8$ a constant introduced to correct the isothermal atmosphere model for muons, and $a=\cos\theta(dE_\mu/dy) = 2.5~{\rm MeV}/{\rm g\:cm}^2$ is the minimum ionizing particle energy loss of the muon along its path through the atmosphere, also assumed constant. 
In Eq.(\ref{eq:sd:Pmu}), Smith and Duller have assumed that the dependence of $P_\mu$ on $y/\cos\theta$ can be substituted for an appropriate average $\langle y/\cos\theta\rangle_{\rm avg} = 100~{\rm g/cm}^2$.
Within the model assumptions, the parent pion energy is given by

\begin{equation}
E_\pi = \frac{1}{r}\left[ E_\mu + a y_0(\sec\theta - 0.100) \right].
\label{eq:sd:Epi}
\end{equation}

In writing equations (\ref{eq:sd:Nmu}), (\ref{eq:sd:Pmu}), and (\ref{eq:sd:Epi}), a number of assumptions and approximations have been made \cite{smith-duller:1959}: pions are produced with a constant average multiplicity, taking a constant fraction of the primary nucleon energy, and are exponentially attenuated with the primary nucleon absorption mean free path $\lambda_P$, which is taken to be approximately equal to $\lambda_\pi$; the forward direction of the primary nucleon is maintained by the produced particles; the curvature of the atmosphere is neglected and an isothermal atmosphere approximation is used including ad-hoc correction factors on the ratio $y_0/\rho_0$ for each particle species; only muons produced by pions are considered although it has been shown that kaon decay contributes in the order of 20\% to the total atmospheric muon flux \cite{rastin:1984}. The model has built into it an omnienergetic angular distribution that to an excellent approximation is $\propto \cos^n\theta$ with $n=2$. 

As noted by the authors of \cite{chatzidakis:2015}, the Smith-Duller model captures correctly the different features of the muon spectrum at various zenith angles, and the approximations and assumptions summarized above have little effect on its accuracy to describe the cosmogenic muon differential intensities at high atmospheric depths, over a wide range of energies.

\begin{figure}[t] 
\centering    
\includegraphics[width=1.0\columnwidth] {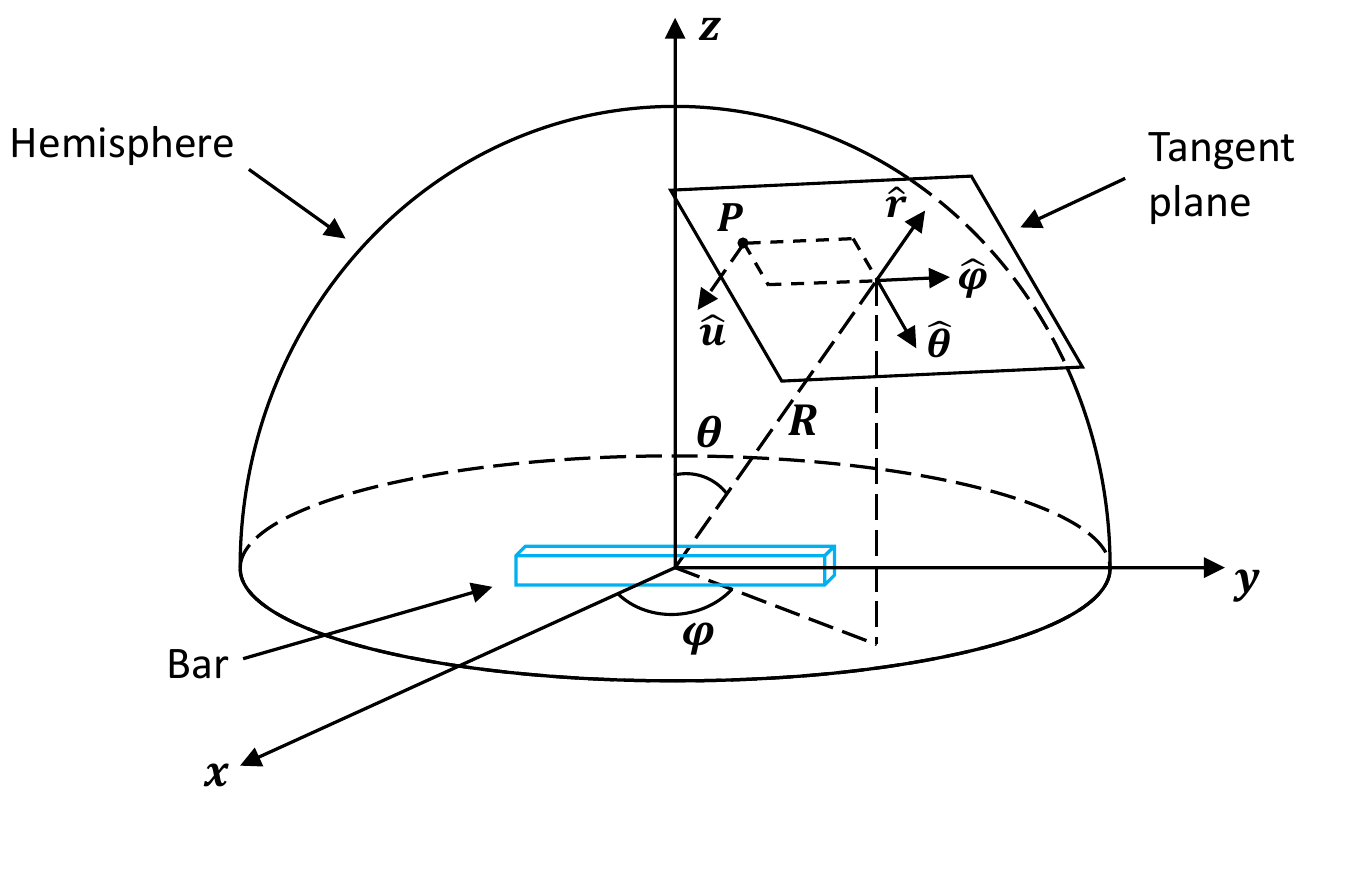}
\caption{\small{Geometry for the muon generator algorithm.}}
\label{fig:gen-algo}
\end{figure}

\begin{figure*}[t] 
\centering    
\includegraphics[width=2\columnwidth] {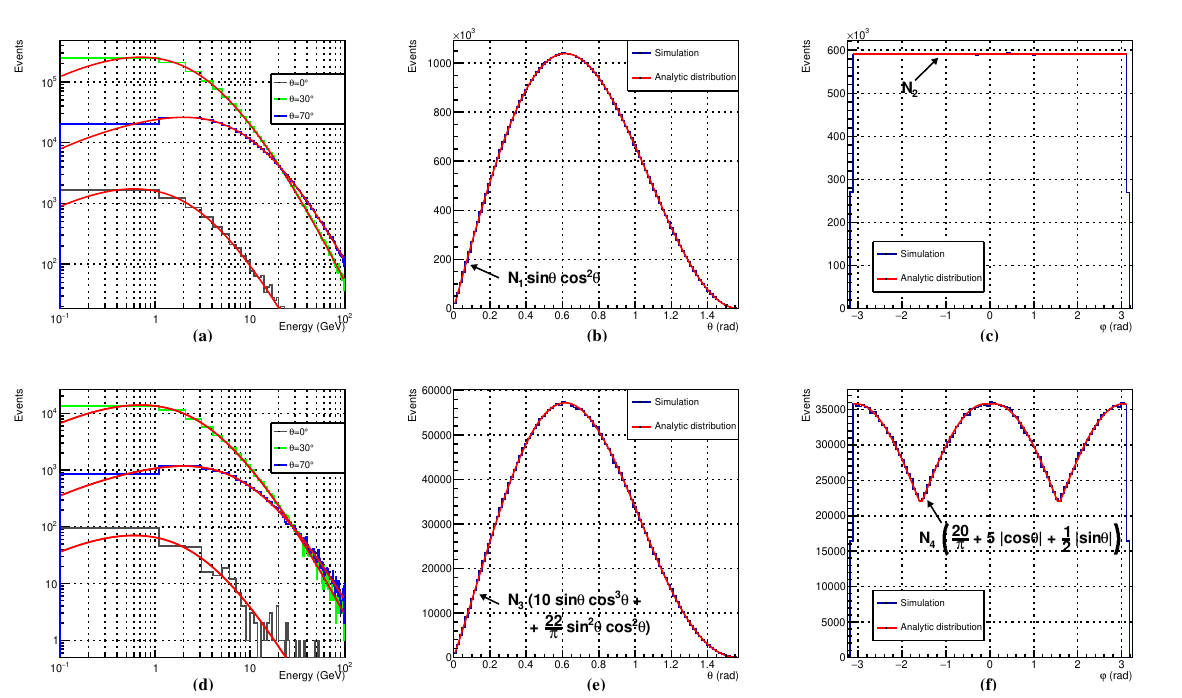}
\caption{\small{Checks of the GEANT4 simulation. The upper plots show the (a) energy, (b) cosine of the zenith angle, and (c) azimuthal angle distributions of the generated muons. The lower plots show the (d) energy, (e) cosine of the zenith angle, and (f) azimuthal angle of the simulated muons entering through any of the faces of the detector. The continuous lines show the expected analytical behavior in each case.}}
\label{fig:sim-checks}
\end{figure*}

\section{\label{sec:generator} Atmospheric muon generator}

Our atmospheric muon generator is based on the geometrical construct shown in Figure~\ref{fig:gen-algo}, and in default mode uses the Smith-Duller model. A hemisphere of radius $R=4.5$~m is placed on the XY plane with the detector resting at the origin of the coordinate system along the Y direction. We begin by drawing the muon zenith and azimuthal angles $\theta$ and $\varphi$ from the omnienergetic angular distribution $dP(\theta,\varphi) = A \cos^2\theta \:d\Omega$, with $\theta$ in [0,$\pi/2$] and $\varphi$ in [0,$2\pi$]. The direction of the generated muon will be given by the vector with Cartesian coordinates in 3D space

\begin{equation}
{\bf \hat u} = -\boldsymbol{\hat{r}} = -(\sin\theta\cos\varphi, \sin\theta\sin\varphi, \cos\theta).
\end{equation}
At the point $R\:\boldsymbol{\hat{r}}$, we then construct a square $\ell\times\ell$ plane ($\ell=1.5$~m), tangent to the sphere. A random point $\boldsymbol{p}$ on this plane is selected by drawing two numbers $a$, and $b$ from a uniform distribution in the interval $[-\ell/2,\ell/2]$ such that

\begin{equation}
\boldsymbol{p} = a\:\boldsymbol{\hat{\theta}} +
b\:\boldsymbol{\hat{\varphi}},
\end{equation}
\noindent
where
\begin{eqnarray}
\boldsymbol{\hat{\theta}} &=&  \boldsymbol{\hat{x}}\;\cos\theta\cos\varphi
                              +\boldsymbol{\hat{y}}\;\cos\theta\sin\varphi 
                              -\boldsymbol{\hat{z}}\;\sin\theta ,    \nonumber \\
\boldsymbol{\hat{\varphi}} &=& -\boldsymbol{\hat{x}}\;\sin\varphi +
                               \boldsymbol{\hat{y}}\;\cos\varphi .
\end{eqnarray}
\noindent
The point $\boldsymbol{p}$ on the tangent plane will be the starting position of the muon, discarding points lying below the $z=0$ plane. The muon energy is drawn from the energy distribution that corresponds to the angle $\theta$ according to Smith-Duller. In this algorithm muons generated at the same $(\theta,\varphi)$ form a shower of parallel rays that rain over the detector's extent, hence the tangent plane dimensions are chosen to cover the detector completely from all possible points on the sphere. 
In principle, the size of the tangent plane should be very large ($l\rightarrow\infty$) in order to allow large angle scatterings that would make a muon originally not aimed at the bar to enter in it. A larger plane would cause the simulation to be less efficient in generating muons that enter the scintillator bar. The agreement between the simulation and analytical results in Figure \ref{fig:sim-checks} suggests that the chosen $l$ is large enough.
In order to explore systematic variations due to the muon flux model, we implemented the capacity to sample from other energy and angular distributions.

\section{\label{sec:simulation} Simulation}

The muon generator algorithm and the detector geometry were implemented in a GEANT4 simulation using the standard electromagnetic physics list, and the QGSP\_BIC\_HP physics list for hadronic interactions. All electromagnetic and particle decay processes, including muon decay, were turned on in the simulation. 
The detector geometry included the plastic scintillator volume with density of 1.032~g/cm$^3$, a 0.08~cm thick layer of PVC around all faces except the small $10~{\rm cm}\times10~{\rm cm}$ faces where the PMTs are. At each end of the bar a coarse model of the PMTs and mu-metal shields was added for completeness. 

The universe containing the hemisphere where muons are generated was filled with air. The Birks energy correction was turned on with a Birks constant of $K_B=1.26\times10^{-2}~{\rm g MeV}^{-1}{\rm cm}^{-2}$, corresponding to PVT material. Only the plastic scintillator volume was defined as a sensitive volume where the deposited energy by traversing particles was collected. 

The effects of PMT response, such as non-linear effects, DAQ noise and saturation, were implemented by means of a convolution of the simulated deposited energy spectrum with a generic Gaussian response function and a non-linear conversion from energy (MeV) to digitizer units (ADC). The parameters of the response function and non-linearity were determined from a fit to experimental spectra, and were later added to the simulation in order to include them in an event by event basis. This was accomplished by adding to the deposited energy of a given event a random Gaussian fluctuation dependent on the value of the deposited energy, and applying the conversion from MeV to ADC according to the non linear model.

The simulation was checked by comparing the energy and angular distributions of the muons entering through the detector faces with theoretical expectations based on the model of the source and the detector geometry. The top plots in Figure~{\ref{fig:sim-checks}} show the distribution of input variables from the muon generator algorithm: (a) energy, (b) zenith angle, and (c) azimuthal angle, compared with the shapes from the the energy spectrum parameterization, and the $dP\propto\cos^2\theta\:d\Omega$ angular distribution. By construction the distribution of the zenith angle is proportional to $\sin\theta\cos^2\theta$, and the distribution of $\varphi$ is uniform. Plots (d), (e), and (f) in the figure show the corresponding distributions for muons that have deposited some energy in the bar. As expected, the energy distribution of muons is mostly unaffected by the requirement that they penetrate the bar, while the angular distributions reflect the fact that the flux of muons entering each face depends on its orientation (horizontal $\it vs$ vertical). For the assumed input angular distribution the zenith and azimuthal angle distributions can be shown to have the exact form
\begin{eqnarray}
\Theta(\theta) &\propto& A \sin\theta\cos^3\theta + B \sin^2\theta\cos^2\theta 
\label{dist-theta}\\
\Phi(\varphi)  &\propto&  C + D |\cos\varphi| + E |\sin\varphi| ,
\label{dist-phi}
\end{eqnarray}
\noindent
where $(A,B,C,D,E) = (10, 22/\pi, 20/\pi,5,1/2)$. This behavior is verified in plots (e) and (f) in Figure~\ref{fig:sim-checks}. For the $\theta$ distribution in Eq.~(\ref{dist-theta}) the first term corresponds to the horizontal face, while the second term correspond to the contribution of all 4 vertical faces. For the $\varphi$ distribution in Eq.~(\ref{dist-phi}) the constant term corresponds to the horizontal face, the $|\cos\varphi|$ term corresponds to the long vertical faces perpendicular to the X-axis, and the $|\sin\varphi|$ term corresponds to the small vertical faces perpendicular to the Y-axis. Note that for our bar the area of the small square faces is one tenth of the long rectangular faces.
 
An approximate geometry of the building was implemented to generate spectra that would be compared to experimental data. The room where the experiment was run has 4.4~m long walls, 4~m in height. Therefore,  the building was modeled as a box with these dimensions, having vertical concrete walls 20~cm thick and a ceiling with 28~cm of concrete. An additional 40~cm layer of brick was added on top of the ceiling to simulate the additional material from the two stories above the actual laboratory. This model of the building fitted within the hemispherical dome used in the muon generator algorithm.

\begin{figure}[t] 
\centering    
\includegraphics[width=1.0\columnwidth] {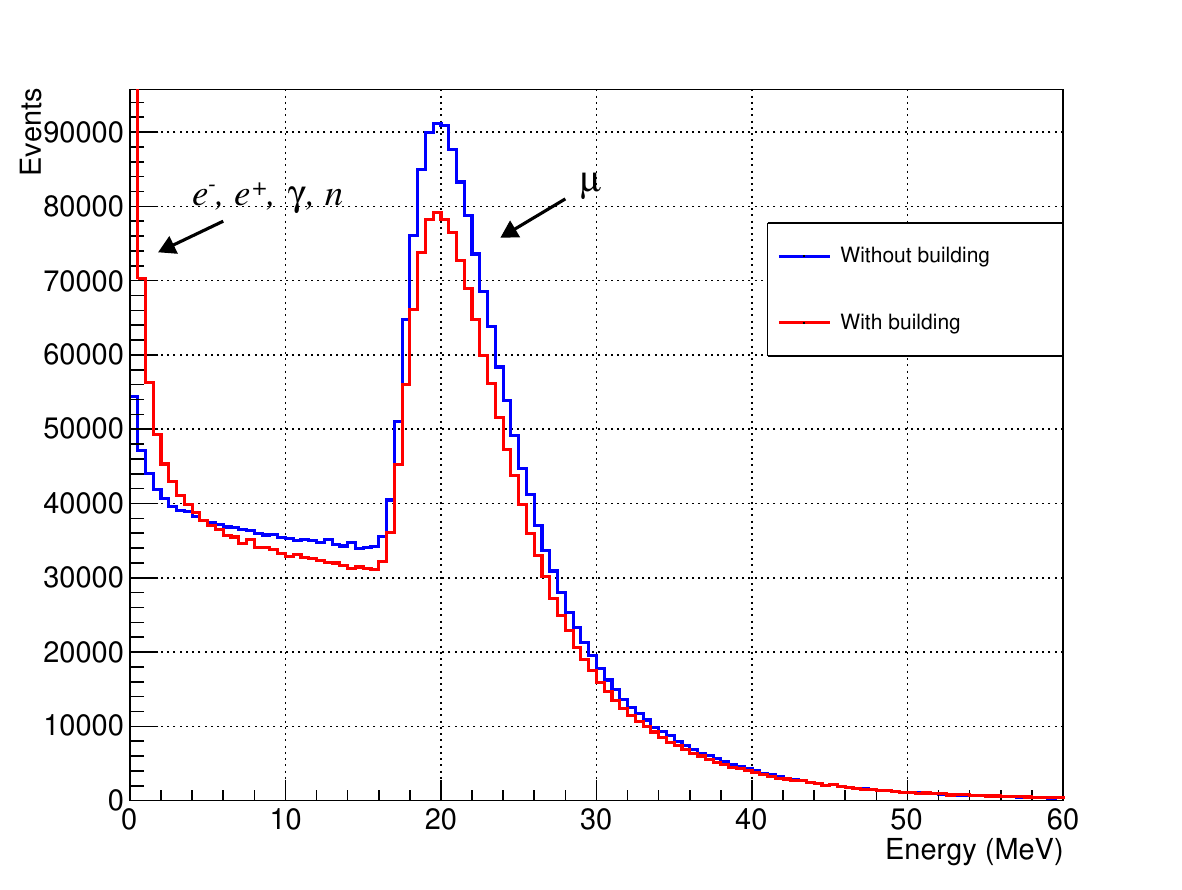}
\caption{\small{Distributions of the energy $E_d$ deposited by muons in the bar according to the simulation with (red) and without (blue) building. The histogram is calculated up to 100 MeV (0.5~MeV/bin), here shown up to 60 MeV.
}}
\label{fig:g4-edep}
\end{figure}

The distribution of the energy deposited by muons in the simulation with and without the building structure are compared in Figure~\ref{fig:g4-edep}. The building has two major effects: {\it i}) reduce the number of muons in the Landau-shaped peak at 20~MeV, corresponding to the energy deposited by vertical minimum ionizing particles (MIPs) traversing $\sim$10~cm of material and depositing $\sim$2~MeV/cm, and {\it ii}) enhance the lower energy part of the spectrum with particles from electromagnetic showers and neutrons.

The simulated muon rate can be calculated by integrating the omnienergetic angular distribution of muons reaching the surface $dN/(dA_{\perp}d\Omega~dt) = I_0 \cos^2\theta$ over the solid angle of the upper hemisphere above the detector, and over the area of the tangent plane from where the muons are originated. The integral vertical muon intensity in the simulation was arbitrarily set to the sea level reference value for muons with energies $>1$~GeV¸ (see Section \ref{sec:intro}) of $I^{\rm sim}_0= 70$~m$^{-2}$s$^{-1}$sr$^{-1}$.
Integration gives the rate

\begin{equation}
\frac{dN}{dt} = \frac{2\pi}{3}\;I^{\rm sim}_0\;\ell^2 = 329.87 \;{\rm s}^{-1} = 1~187~532~{\rm h}^{-1}.
\label{eq:rate-sim}
\end{equation}


A total of 50,000,000 muons were simulated, corresponding to a detector exposure time of $T_{\rm sim}= 42.10$~h. 
For angular distributions with $n\ne 2$ the calculation in Eq.~(\ref{eq:rate-sim}) is done changing the $3\rightarrow n+1$ in the denominator. 
According to the simulation and the fitted parameters of the detector response, only muons with $E\gtrsim 110$~MeV produce pulses with integrated charge $>100$~ADC ($E_d>2.6$~MeV).

\begin{figure}[t] 
\centering    
\includegraphics[width=1\columnwidth] {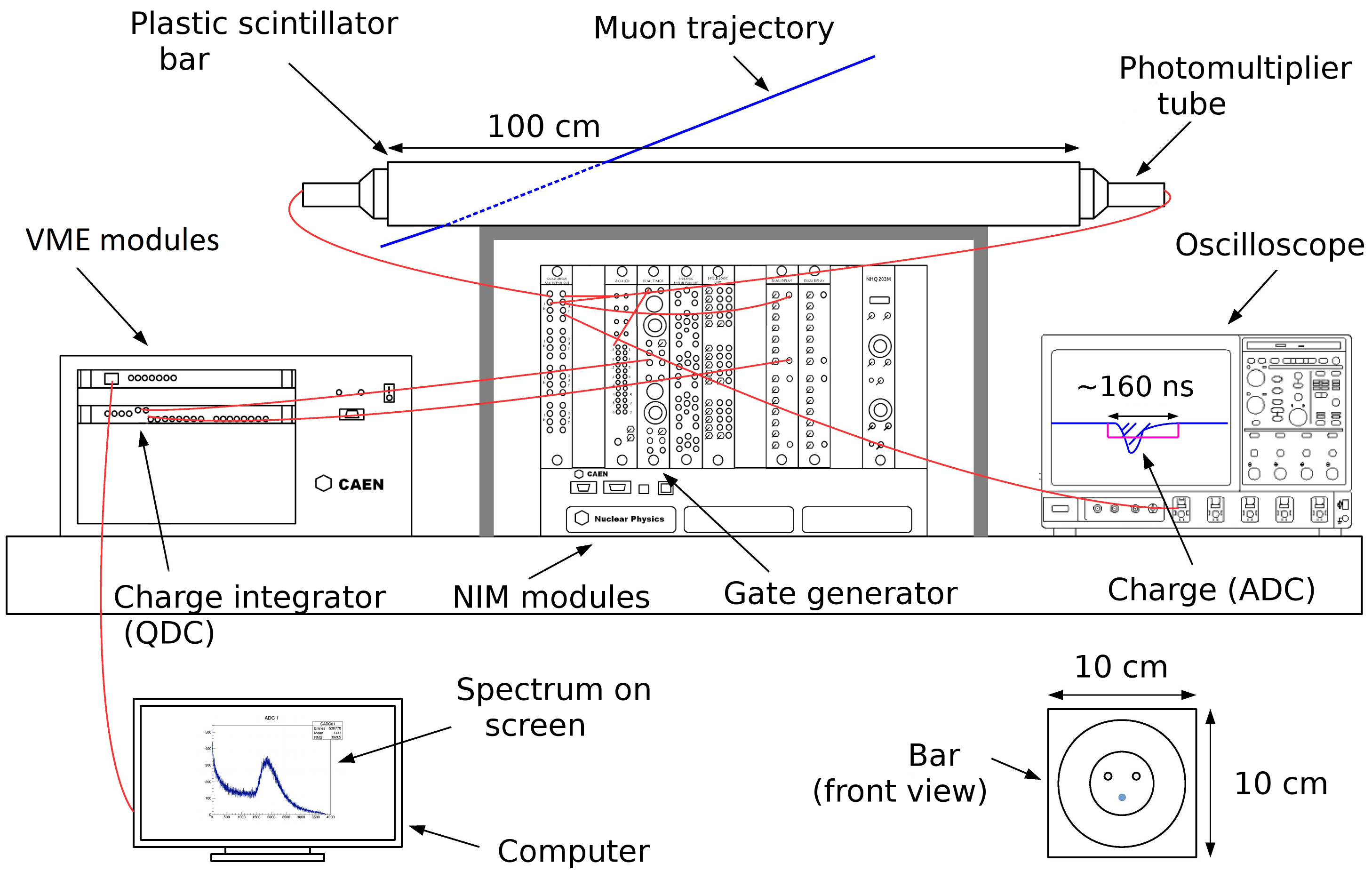}
\caption{\small{Schematic of the experimental setup.}}
\label{fig:exp-schematic}
\end{figure}

\section{Experiment}
\label{sec:exper}

A schematic of the experiment is shown in Figure \ref{fig:exp-schematic}. It consists of a PVT-based plastic scintillator bar (Rexon RP-408) with dimensions $100~{\rm cm}\times10~{\rm cm}\times10~{\rm cm}$ with a 3" photomultiplier (PMT) tube (Rexon-R1200P) encased in a mu-metal shield, and attached to each end. The bar is wrapped in a reflective aluminum paper cover, an intermediate layer of threaded tape, and an outer layer of black PVC tape.
The analog sum of the two PMT signals is processed in a self-trigger mode. One copy of the summed pulse is sent to a discriminator module whose output triggers the opening a 160~ns wide pulse in a gate generator, which is then input as the ``Gate" signal of a CAEN V965 QDC charge integrator module. A second copy of the analog sum of the PMT signals, produced simultaneously with the first one, is sent through a delay module adjusted to fit the signal in the integration gate, and then sent into one of the QDC input channels. A CAEN V1718 USB bridge is used to interface with the DAQ computer where the spectrum of the integrated charge of the summed pulse is accumulated.
With the discriminator level set at $50$~mV, the trigger rate averaged $\sim$410~Hz. At this low rate the probability to have two events in the same trigger window is $<10^{-6}$, and the efficiency of the DAQ was measured to be $\epsilon=0.998\pm 0.003$~\% for pulses with frequencies below 1~kHz. This efficiency was measured sending square pulses to the DAQ  which were digitized with pulse heights corresponding to ADC values of 2200, {\it i.e.}  slightly above the peak of the muon spectrum.

The experiment was run in the interior of the Detectors Laboratory at ICN-UNAM in Mexico City (19.33$^\circ$ latitude, 99.19$^\circ$ longitude, 2,268~m sea level altitude, geomagnetic cutoff 8.2~GV \cite{vargas:2013,smart:2009}), under an effective coverage of $\sim$28~cm of concrete and $40$~cm of brick from the ceiling, and $\sim$20~cm of concrete from the walls.
Spectra were acquired over 5~hr with the detector oriented $\sim$10$^\circ$ from the N-S direction (clock wise looking from above). \\

\section{Fit to data}
\label{sec:fits}

\begin{table*}[t]
\caption{\small Parameters of the convolution function, Eq.(\ref{eq:conv}), fitted to the experimental spectrum for different flux models (see text). The last row shows the extracted value of the muon vertical intensity. The errors are from the fit statistics.}
\centering
\begin{tabular*}{\textwidth}{c @{\extracolsep{\fill}} cccccc} \hline
Parameter & Smith \& Duller & EXPACS ($\cos^2\theta$) & EXPACS ($f(\theta)$) & Reyna & Units\\ \hline
$f$ & $4.51\pm 0.18$ & $4.39\pm 0.18$ & $5.20\pm 0.15$ & $4.55\pm 0.17$ &  $(\times 10^{-2})$\\
$N_\mu$ & $5.36\pm0.03$ & $5.29\pm0.03$ & $5.14\pm0.03$ & $5.34\pm0.03$ & $(\times10^5)$\\
$N_{b_1}$ & $0.88\pm 0.30$ & $0.91\pm 0.26$ & $0.97\pm 0.06$ & $1.00\pm 0.10$ & $(\times 10^5)$\\
$\varepsilon_1$ & $2.20\pm 0.42$ & $2.15\pm 0.39$ & $2.08\pm 0.15$ & $2.30\pm 0.19$ & MeV\\
$N_{b_2}$ & $4.10\pm 3.15$ & $4.22\pm 2.75$ & $5.27\pm 0.59$ & $3.06\pm 1.02$ & $(\times 10^4)$ \\
$\varepsilon_2$ & $4.97\pm 2.01$ & $5.18\pm 2.02$ & $7.28\pm 1.11$ & $6.61\pm 2.01$ & MeV\\
$a_0$ & $-2.175$ & $-2.175$ & $-2.175$ & $-2.175$ &  $(\times 10^2)$~ADC\\
$a_1$ & $1.24\pm 0.004$ & $1.24\pm 0.004$ & $1.26\pm 0.006$ & $1.24\pm 0.005$ &  $(\times 10^2)$~ADC/MeV\\
$a_2$ & $1.07\pm 0.02$ & $1.07\pm 0.02$ & $1.12\pm 0.03$ & $1.03\pm 0.02$ &  $(\times 10^{-2})~$MeV$^{-1}$\\
&&&&&\\
$I_0$ & $107.2\pm 0.51$ & $112.6\pm 0.60$ & $110.0\pm 0.74$ & $103.7\pm 0.61$ & m$^{-2}$ s$^{-1}$ sr$^{-1}$\\ \hline
$\chi^2$/ndf & 330.3/332 & 326.4/332 & 348.5/332 & 329.7/332 & \\
\end{tabular*}
\label{tab:fitpars}
\end{table*}

A convolution function combining the predicted deposited energy spectrum from the Geant4 simulation, and the effects of random fluctuations and non linearity in the detector energy response was constructed to fit it to the experimental spectrum. 
The function has nine free parameters: The muon spectrum normalization ($N_\mu$), the normalizations of the low energy exponentially decaying backgrounds ($N_{b_1}$ and $N_{b_2}$), the exponential decay constants of the low energy backgrounds ($\varepsilon_1$ and $\varepsilon_2$), the fractional energy resolution at 20~MeV of deposited energy ($f$), and the three parameters of the non linear model for the detector response ($a_0$, $a_1$ and $a_2$). 
The fit function is given in Eq.(\ref{eq:conv})

\begin{widetext}
\begin{equation}
{\cal F}(E_a) = \frac{1}{(dE_a/dE_v)} \int_{0}^{E_{\rm max}}
\left( 
  N_\mu \; F_\mu(E_d) 
  + N_{b_1} \; (1/\varepsilon_1) e^{-E_d/\varepsilon_1} 
  + N_{b_2} \; (1/\varepsilon_2) e^{-E_d/\varepsilon_2}
\right) \;
\frac{1}{\sqrt{2\pi\sigma^2}}e^{-\frac{(E_v-E_d)^2}{2\sigma^2}} \;dE_d
\label{eq:conv}
\end{equation}
\end{widetext}
%
%
%

%
\begin{figure}[t] 
\centering
\includegraphics[width=1.0\columnwidth] {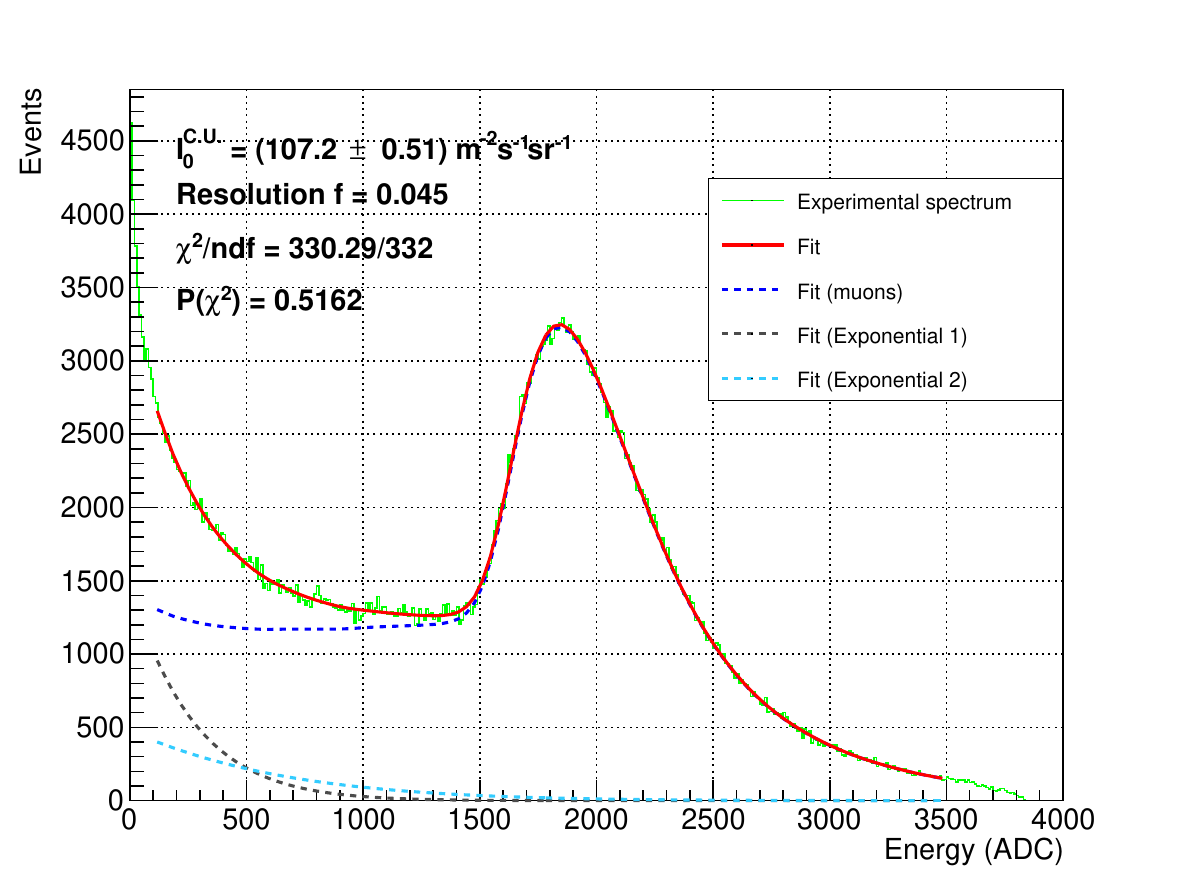}
\caption{\small{Experimental spectrum (green) fitted with the convolution integral of Eq.(\ref{eq:conv}) in the range of 100-3500 ADC (red). The cosmogenic muon component is shown in blue.}}
\label{fig:fitted-spectrum}
\end{figure}

\noindent In the convolution function, Eq.(\ref{eq:conv}), $E_a$ is the energy in digitizer units (ADC), $E_d$ is the energy, in MeV, deposited by the particle in the sensitive volume of plastic scintillator according to the Geant4 simulation, and $E_v$ is the ``visible" energy, in MeV, resulting from an overall Gaussian response at deposited energy $E_d$. All non-linear and saturation effects, independently of their origin (detector or electronics) are accounted for by means of the conversion of the spectrum from units of MeV to ADC, effected by the factor $(dE_a/dE_v)^{-1}$. We also use the definitions

\begin{equation}
\sigma = f \sqrt{E_0 E_d} \;, \;\;E_0=20~{\rm MeV},
\label{eq:def-sigma}
\end{equation}
\noindent and
\begin{equation}
\frac{dE_a}{dE_v} = \frac{a_1}{(1+a_2 E_v)^2} , \;\;\;
E_v = \frac{E_a-a_0}{a_1+a_0a_2-a_2E_a} .
\label{eq:change-units}
\end{equation}

\noindent
$F_\mu(E_d)=\left(1/{\cal N}^{\rm sim}_\mu \right)d{\cal N_\mu}/dE_d$, is the unit-normalized deposited energy spectrum obtained by dividing the Geant4 simulated spectrum in Figure~\ref{fig:g4-edep} (with building) by the bin width (0.5~MeV/bin) and the total number of entries in the simulated histogram; its integral from $0$ to $E_{\rm max}=100~{\rm MeV}$ is by definition set to one. The shape of the low energy part of the spectrum, due to neutrons and electromagnetic showers, is not included in the simulation and is modeled as two simple exponentials $(1/\varepsilon_{1,2}) e^{-E_d/\varepsilon_{1,2}}$ with decaying constants $\varepsilon_{1,2}$, in MeV. Both spectral shapes (muons and low-$E_d$ backgrounds) are treated on equal footing with respect to the modelled detector response.

The function ${\cal F}(E_a)=d{\cal N}/dE_a$, gives the number of events per visible energy interval (events per ADC), and was fitted to the experimental spectrum in the interval from 100 to 3500 ADC, as shown in Figure~\ref{fig:fitted-spectrum}. In the fit the constant $a_0$ from the non-linearity model was kept fixed to a value consistent with the DAQ pedestal.
Given the assumed dependence of the Gaussian width in Eq.~(\ref{eq:def-sigma}), for pulses with $E_a>100$~ADC the effect of truncating the Gaussian response function at $E_d=0$ in the convolution integral is negligible.
Table \ref{tab:fitpars} shows the values of the fitted parameters and their uncertainties. 

The position and shape of the peak in the spectrum are determined by the last four parameters in Table \ref{tab:fitpars}.  The overall energy response is consistent with a $\sim$5\% energy resolution  at 20~MeV (parameter $f$). Eq.~(\ref{eq:change-units}) implies that
\begin{equation}
E_{\rm ADC} = a_0 + \frac{a_1 E_v}{1+a_2E_v}, 
\label{eq:change-units2}
\end{equation}

\noindent
from where we see that parameter $a_0=-217.5$~ADC (fixed) indicates the position of the zero energy in the histogram scale (pedestal). Parameter $a_1\approx 124$~ADC/MeV gives the linear part of the conversion factor from energy to digitizer units, and parameter $a_2\approx 1$~\%/MeV represents the non-linearity in the detector response, which is significant ($>10\%$) for energies above 10~MeV, and hence for most of the spectrum. This large non-linearity was expected, given that the PMTs were operated near their maximum recommended voltages in order to make the flat part of the muon spectrum more clearly visible.

\begin{figure}[t] 
\centering
\includegraphics[width=1.0\columnwidth] {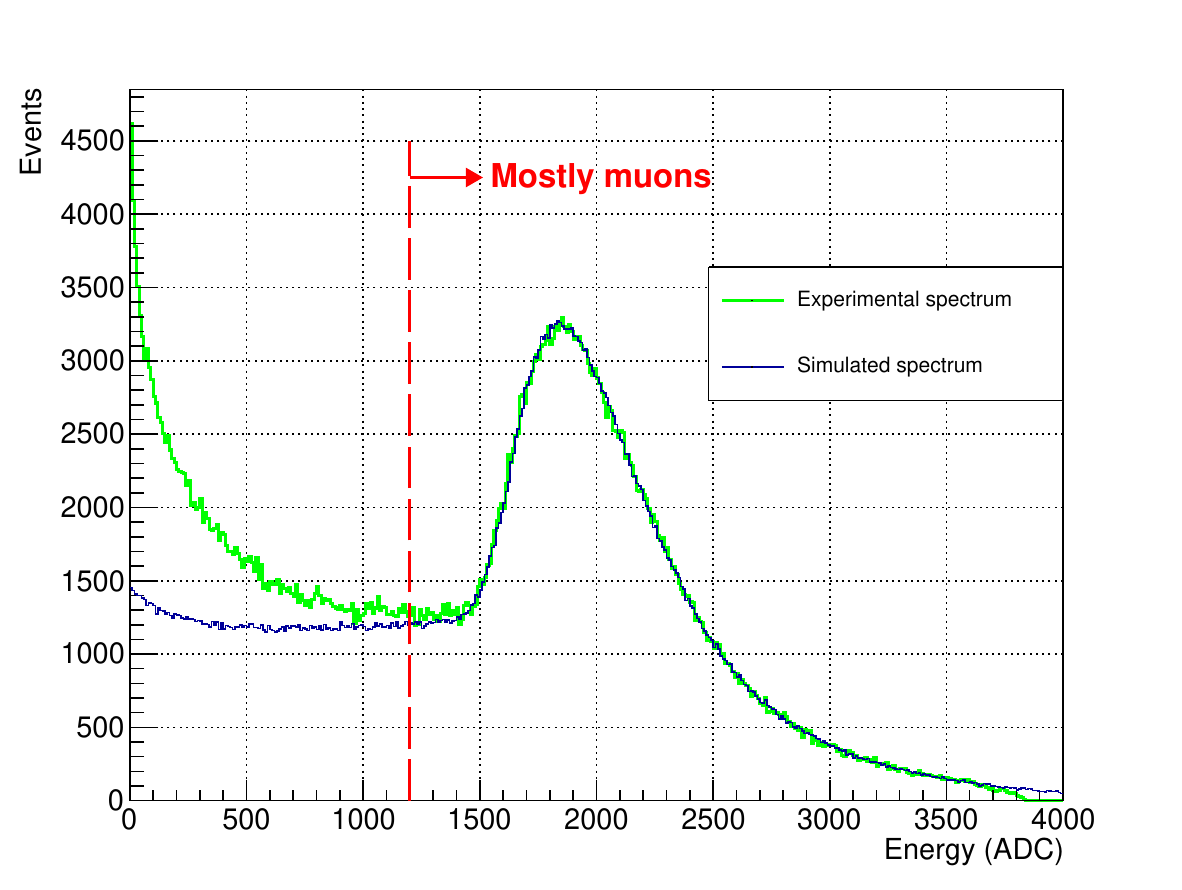}
\caption{\small{Simulated spectrum with resolution and non-linearity effects included event by event (blue) compared to experimental spectrum (green). The simulation used the parameters $f$, $a_0$, $a_1$ and $a_2$ from Table \ref{tab:fitpars}}.}
\label{fig:full-sim-spectrum}
\end{figure}

As a check that the convolution function, Eq.~(\ref{eq:conv}), was representing the physical effects of fluctuations and non-linear effects properly, we completed the Geant4 simulation by adding these effects event-by-event. Given the deposited energy $E_d$ in an event, a Gaussian fluctuation with $\sigma$ given by Eq.~(\ref{eq:def-sigma}) was added to it, $E_d'=E_d+\delta E_d$, and the energy in digitizer units was assigned using Eq.~(\ref{eq:change-units2}). Figure \ref{fig:full-sim-spectrum} shows the comparison of the simulated spectrum incorporating the resolution and non-linearity in an event-by event basis, demonstrating the equivalence with the convolution approach.

\section{\label{sec:muon-int}Integral vertical muon intensity and angular distribution}

The muon normalization parameter ($N_\mu$) in Table \ref{tab:fitpars} is proportional to the integral vertical muon intensity at the experiment's location ($I_0^{\rm lab}$), and to the running time ($T$), therefore

\begin{equation}
I_0^{\rm lab} = I_0^{\rm sim} 
\left( \frac{N_\mu}{{\cal N}_\mu^{\rm sim}} \right) 
\left( \frac{T_{\rm sim}}{T} \right) 
\times \left( \frac{1}{\epsilon} \right) ,
\label{eq:I0}
\end{equation}

\noindent
where $I_0^{\rm sim}$ is the integral vertical muon intensity used in the muon generator in Section \ref{sec:simulation}, $T_{\rm sim}$ is the simulation time, and $\epsilon$ is an effective detection efficiency, which we took as the DAQ efficiency of Section \ref{sec:exper}. Due to being a small correction, in what follows we approximated $\epsilon \approx 1$, but considered its effect in the systematic uncertainty for $I_0$. For 5~hr long exposures, this yields a measurement of the integral vertical muon intensity of  $I_0 = (107.2 \pm 0.51)~{\rm m}^{-2}{\rm s}^{-1}{\rm sr}^{-1}$ (statistical error only) using our benchmark Smith-Duller model of the flux, which we will consider our nominal result for $n=2$.
%
\begin{table}[b]
\caption{\small Summary of systematic errors for fixed $n=2$. The percent error is relative to our nominal value (Smith-Duller) of 
$I_0 = 107.2~ {\rm m}^{-2}{\rm s}^{-2}{\rm sr}^{-1}$.}
\begin{tabular*}{\columnwidth}{@{\extracolsep{\fill}} lcc} \hline
Source & Error (m$^{-2}$s$^{-1}$sr$^{-1}$) & (\%)\\ \hline
Flux model                     &  5.0   & 4.6 \\
Building coverage ($\pm 50\%$) &  2.8   & 2.6\\
DAQ efficiency                 &  0.2   & 0.2 \\ \hline
Total                          &  5.8   & 5.4 \\ \hline
\end{tabular*}
\label{tab:errors}
\end{table}

As an attempt to assess the systematic uncertainty in $I_0$ arising from the modeling of the atmospheric muon flux we considered three alternative models. 
The first one used the energy distribution calculated with the EXPACS \cite{expacs1:2016, expacs2:2015} tool at the geographic coordinates of Mexico City, but forcing a $\cos^2\theta$ angular distribution.
A second one used both, the energy and angular distributions as predicted by EXPACS, forcing the angular distribution to have the same normalization as the $\cos^2\theta$ case.
The third model considered was that of Reyna \cite{reyna:2006}, which also gives an angular distribution proportional to $\cos^2\theta$. 

Table \ref{tab:fitpars} shows the result of the fits of the convolution function Eq.(\ref{eq:conv}) with each of the models. The extracted parameters are in general very consistent, and are in reasonably good agreement with expectations derived from older measurements at varying altitudes, for locations away from strong geomagnetic anomalies \cite{pdg:2018}.
Table \ref{tab:errors} summarizes the systematic uncertainties that were considered for the measurement of $I_0$. The effect of the flux model on $I_0$ with respect to the nominal is 5.4\% (RMS about the nominal). In addition, conservative variations on the effective building thickness in the simulation ($\pm 50\%$) were studied and seen to produce a 2.6\% variation on $I_0$. Including the DAQ inefficiency uncertainty of 0.2\%, we estimate a total systematic error of 5.4\%.
Considering these systematic variations, for fixed $n=2$, the nominal integral vertical muon intensity is $I_0=(107.2\pm 0.51 ({\rm stat})\pm 5.8 ({\rm syst}))~{\rm m}^{-2}{\rm s}^{-1}{\rm sr}^{-1}$.

\begin{figure}[t] 
\centering
\includegraphics[width=1.0\columnwidth] {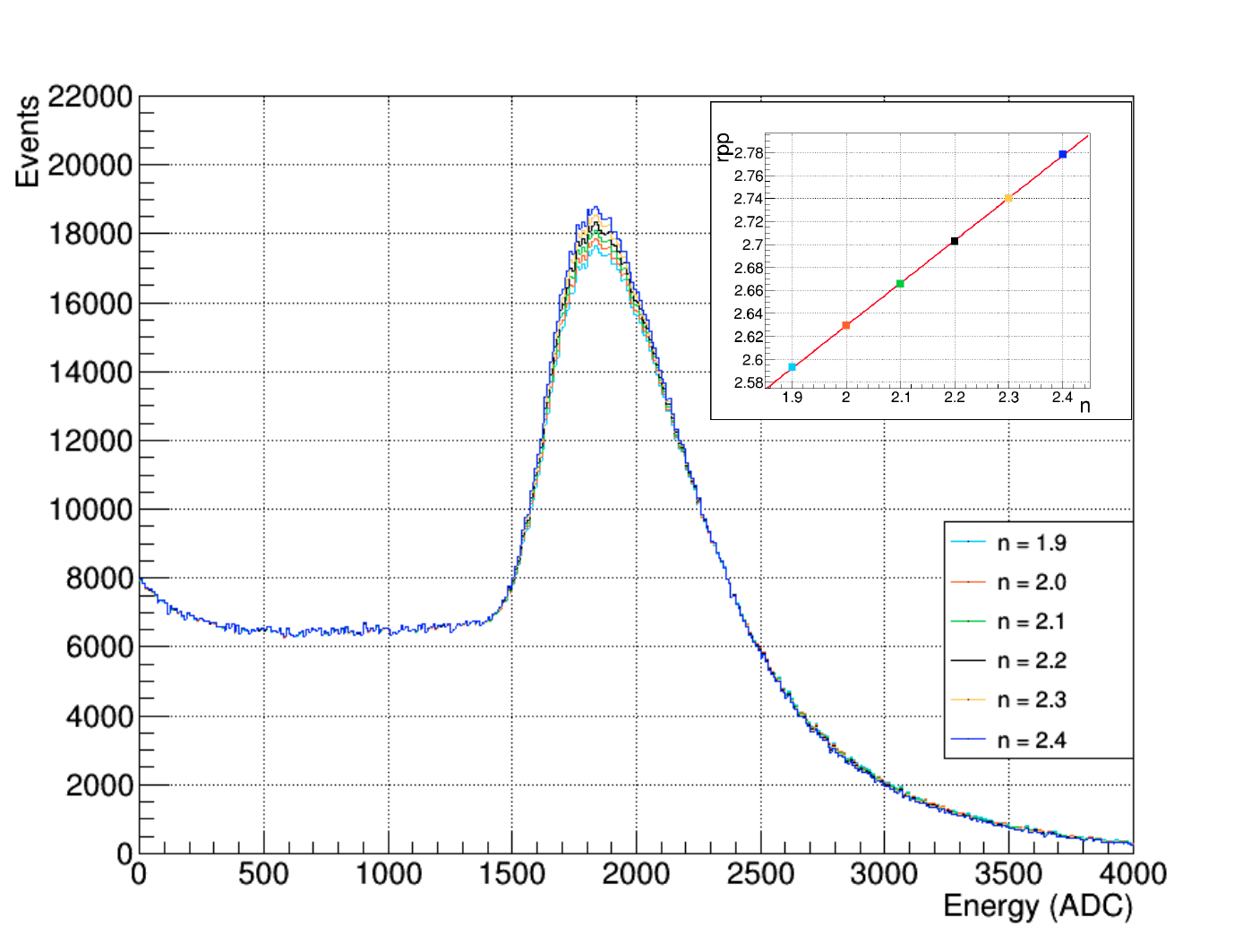}
\caption{\small{Simulated muon energy spectra for different values of the exponent of the angular distribution. The spectra have been arbitrarily normalized to match in the region of the plateau ($<1500$~ADC). The insert shows that the peak-to-plateau ratio ($rpp$) increases linearly with $n$.}}
\label{fig:ratiopp}
\end{figure}

\begin{figure}[!b] 
\centering
\includegraphics[width=1.0\columnwidth] {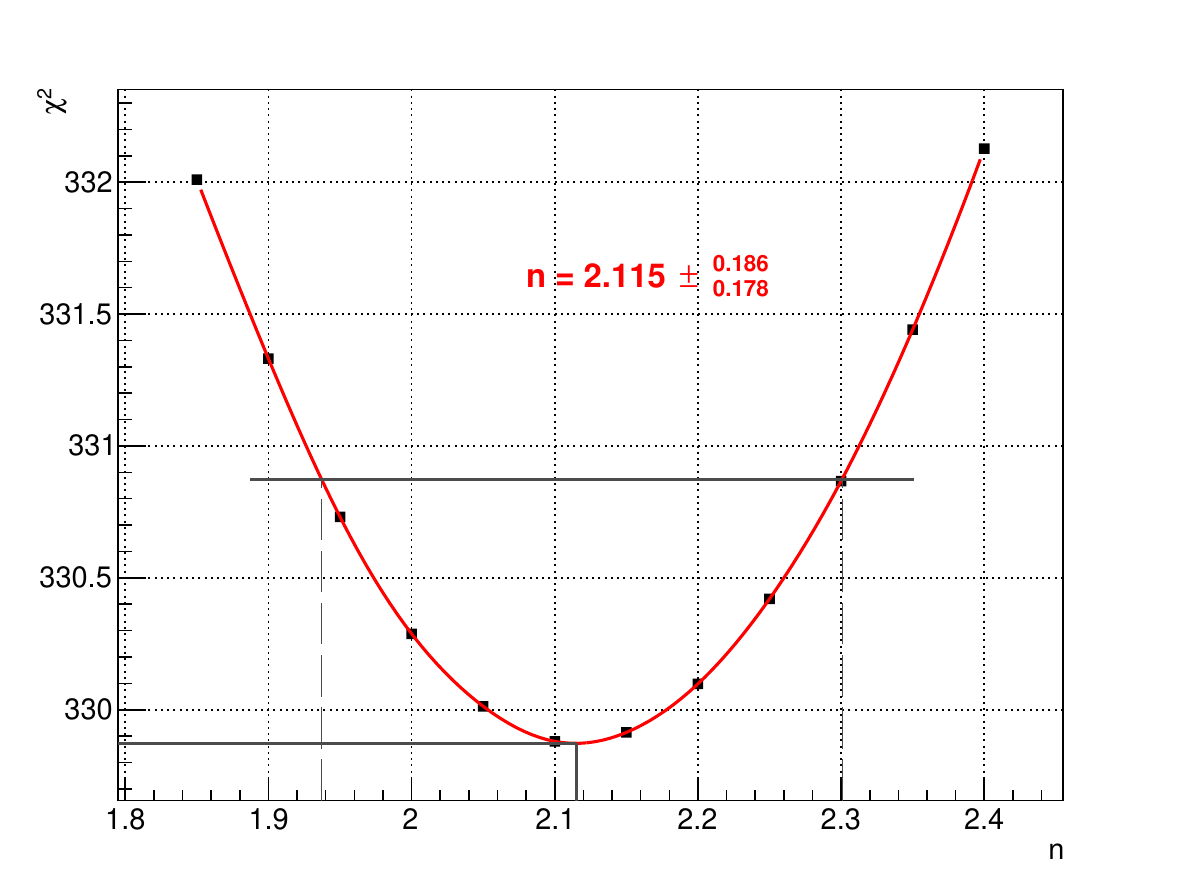}
\caption{\small{The $\chi^2$ minimum obtained from fits of the convolution function, Eq.(\ref{eq:conv}), the experimental data, using different values of $n$ in the simulated deposited energy spectrum $F_\mu(E_d)$. The minimum and 1-$\sigma$ interval ($\Delta \chi^2=1$) represent the marginalization over the 8 free parameters in the fit.}}
\label{fig:chisquared-n}
\end{figure}

Although an angular distribution with $n=2$ provides a very good description of the observed spectrum in Figure \ref{fig:fitted-spectrum}, the data showed a preference for a slightly larger value. 
The preferred exponent in the muon angular distribution was measured by constructing a set of simulations with varying values of $n$ and performing, for each one, a fit for all 8 free parameters in Eq.(\ref{eq:conv}). 
The simulations were constructed by applying a reweighing factor $\cos^n\theta/\cos^2\theta$ to each event initially generated with the Smith-Duller model. Adding this weight to the events modifies the relative height of the Landau-shape peak and the flat portion to its left (the plateau) in the deposited energy distribution. The effect of varying $n$ in the observed energy distribution (in ADC units) is shown in Figure~\ref{fig:ratiopp}. A larger $n$ produces a spectrum with more vertical muons, hence adding to the population under the peak, relative to those entering at wider zenith angles, which contribute to the plateau (corner clipping muons). 
The magnitude of this effect is particular to the geometry of the setup used in this work. If the detector were, for example, a very thick slab of scintillator, the corner clipping muons could become comparable or even dominate over the Landau peak, diluting the effect. Also, for more general detector geometries, vertical muons could also produce corner (or even round border) clipping tracks, changing the trend significantly.

The insert in the figure shows the ratio of the peak height to the plateau height at 600 ADC (ratio peak-to-plateau, rpp), as a function of $n$, demonstrating the sensitivity of the spectral shape to the angular distribution exponent.

In Figure \ref{fig:chisquared-n} we plot the $\chi^2$ minimum from each fit of the convolution function as a function of $n$. The minimum $\chi^2$ over all the fits, as well as the 1$\sigma$ interval ($\Delta \chi^2=1$) are shown. The measured value of the exponent of the angular distribution exponent obtained in this way is $n=2.12\pm ^{+0.19}_{-0.18}$ (statistical error only). The same exercise was repeated using the EXPACS and Reyna flux models, and varying the building thickness in $\pm50\%$, resulting on an RMS variation of 5\% on the value of $n$, comparable to the statistical error. Combining both effects we report a measured value of $n=(2.12\pm0.19 ({\rm stat})\pm0.11({\rm syst}))$, for this initial orientation of the scintillator bar (see end of Section \ref{sec:exper}).


At the preferred value of $n$, the simulation time $T_{\rm sim}$ must be recalculated, as noted near the end of Section \ref{sec:simulation}. Taking this into account, the measured integral vertical intensity is changed to $I_0=(114.9\pm 12.9 ({\rm stat}) \pm 6.2({\rm syst}))~{\rm m}^{-2}{\rm s}^{-1}{\rm sr}^{-1}$, where the large statistical uncertainty ($\sim 11\%$) arises from the variation of $n$ within the 1$\sigma$ interval derived from Figure \ref{fig:chisquared-n}, and calculating $I_0$ with the corrected $T_{\rm sim}$ at each value. Within errors, this is consistent with the measurement reported assuming $n=2$.  Both measurements are in reasonable agreement with expectations for the $\mu^+ + \mu^-$ flux at this altitude (see for example Figure 29.4 in \cite{pdg:2018}), absent strong geomagnetic anomalies.

\section{\label{sec:closure-tests} Closure tests }

In order to assess the robustness of the fitting procedure and the validity of the estimated uncertainties on $I_0$ and $n$ we performed fits to simulated spectra with input parameters $I_0=70$ ${\rm m}^{-2}{\rm s}^{-1}{\rm sr}^{-1}$, and $n=2$ (the rest of the parameters in the simulation were set to those in the second column of Table~\ref{tab:fitpars} in two scenarios.

Scenario I considered a simulation with only muons impinging on the detector without the overburden of the building structure and without exponentially decaying background components ($N_{b_1}=N_{b_2}=0$). The deposited energy spectrum used in this simulation corresponds to the Smith-Duller model and is shown as the blue line in Figure \ref{fig:g4-edep}. The fit to the convolution function was performed letting vary the parameters $f$, $N_\mu$, $a_1$, $a_2$, and also $n$ in the fashion described in Section \ref{sec:muon-int}. The fit returned the values $I_0 = (70.1\pm3.6)$ ${\rm m}^{-2}{\rm s}^{-1}{\rm sr}^{-1}$, and $n=(2.01\pm0.08)$ (statistical errors only), which are consistent with the expected input parameters within errors.

It is interesting to see what is the result of fitting the data spectrum in Figure \ref{fig:fitted-spectrum} with the model without the building coverage, but turning on the parameters describing the low energy backgrounds. The results are $I_0 = (101.9^{+15.3}_{-12.6})$ ${\rm m}^{-2}{\rm s}^{-1}{\rm sr}^{-1}$ and $n=(1.95^{+0.25}_{-0.21})$ (statistical errors only), which are $\sim11\%$ ($I_0$) and $\sim8\%$ ($n$) lower than the fitted values considering the building overburden, but are still within one standard deviation from one another.

\begin{figure}[t]
    \centering
    \scalebox{0.39}{\includegraphics{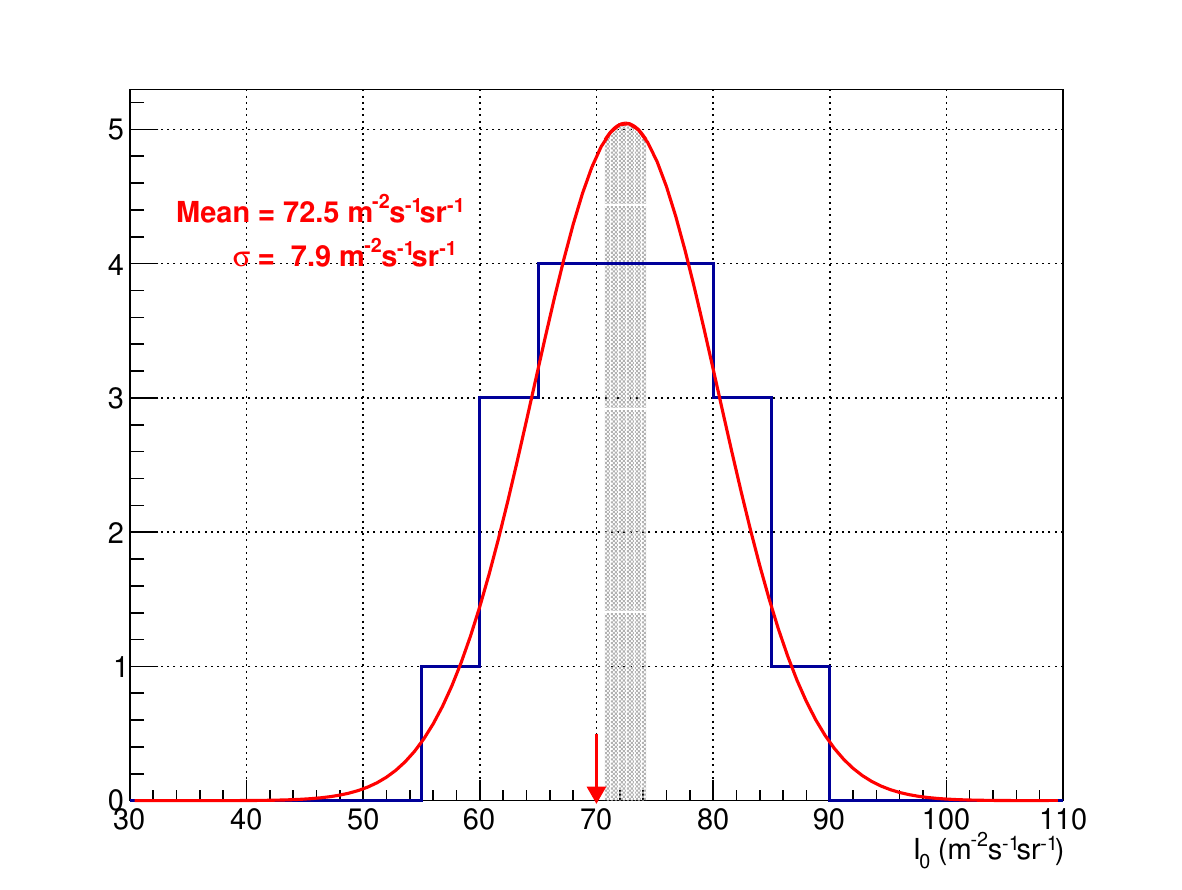}}
    \scalebox{0.39}{\includegraphics{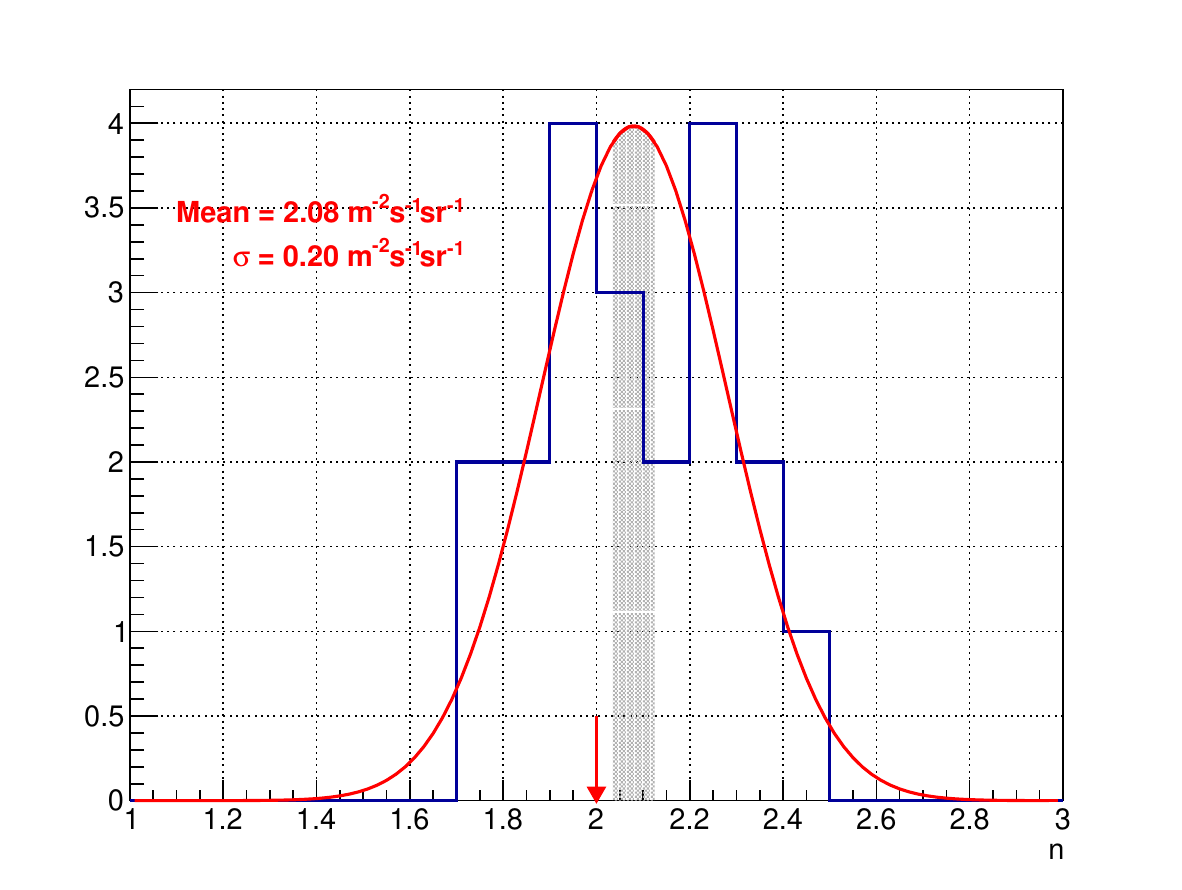}}
    \caption{\small Distributions of fit results for $I_0$ (top) and $n$ (bottom) from the closure test in the Scenario II (see text). The arrows show the location of the input parameters ($I_0 = 70$ ${\rm m}^{-2}{\rm s}^{-1}{\rm sr}^{-1}$, and $n= 2$) and the shaded regions indicate the 1$\sigma$ error on the sample means. The red curves are Gaussian fits to the sampled distributions.}
    \label{fig:closure-with-building}
\end{figure}

Scenario II considered the full simulation with building overburden and two exponentially decaying backgrounds at low energies as applied to the experimental spectrum in Section \ref{sec:fits}.
The fit was performed as in scenario I but letting vary in addition the parameters $N_{b_1}$, $\varepsilon_1$, $N_{b_2}$ and $\varepsilon_2$. Figure~\ref{fig:closure-with-building} shows the distribution of fitted values of $I_0$ and $n$ from a collection of 20 statistically independent simulated spectra. These distributions are well described by Gaussians with the following means and standard deviations:

\vspace{0.3cm}
\begin{tabular}{crl}
$I_0$: & mean &= 72.5 ${\rm m}^{-2}{\rm s}^{-1}{\rm sr}^{-1}$   \\
    & $\sigma$ &= \:\: 7.9 ${\rm m}^{-2}{\rm s}^{-1}{\rm sr}^{-1}$   \vspace{0.3cm} \\ 
$n$: & mean &= 2.08 \hspace{1cm}     \\
    & $\sigma$ &= 0.20 \hspace{1cm}
\end{tabular}
\vspace{0.3cm}

\noindent
The means of the distributions, with their $1\sigma$ error interval shown by the gray bands, are consistent with the input parameters, shown by the red arrows, with a small ($\sim4\%$) offset toward larger vales that is not unexpected from the relatively small sample size. May this systematic offset persist in a closure test with higher statistics, it could be corrected for or added to the error budget. For the time being we will not include it in our results.
The means and standard deviations of these distributions indicate that the statistical uncertainty expected on $I_0$ and $n$ from a fit to any given spectrum should be of the order of 10-11\%. This is consistent with the errors estimated from the fit to the data spectrum in Figure  \ref{fig:fitted-spectrum}.

In all the fits involved in these tests the other fitted parameters, including those related to the detector response $f$, $a_1$, and $a_2$ were allowed to vary freely ($a_0$ is a fixed offset of the energy scale set in the DAQ and was not allowed to float). The fitted values have been carefully inspected and found to lie within reasonable fluctuations from the expected ones.

\begin{figure}[t]
    \centering
    \includegraphics[width=1.0\columnwidth]{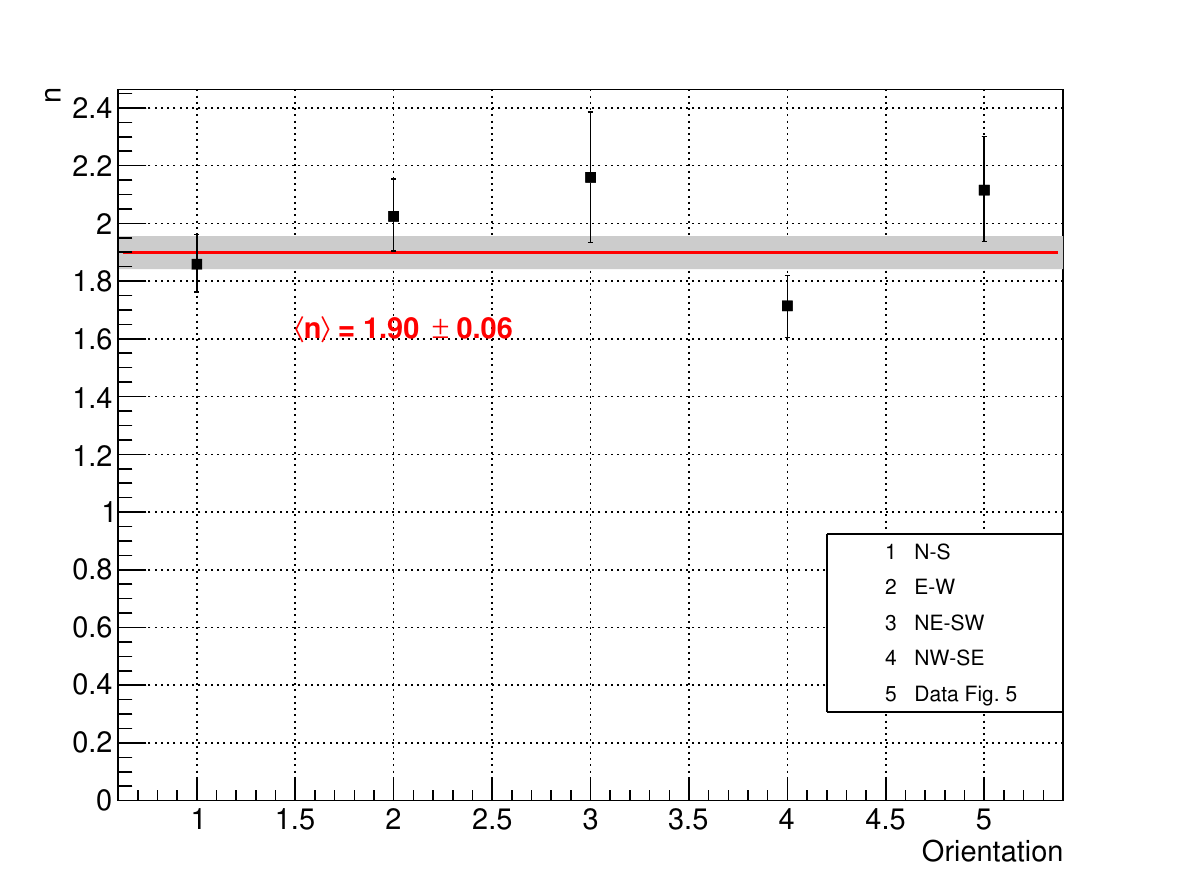}
    \includegraphics[width=1.0\columnwidth]{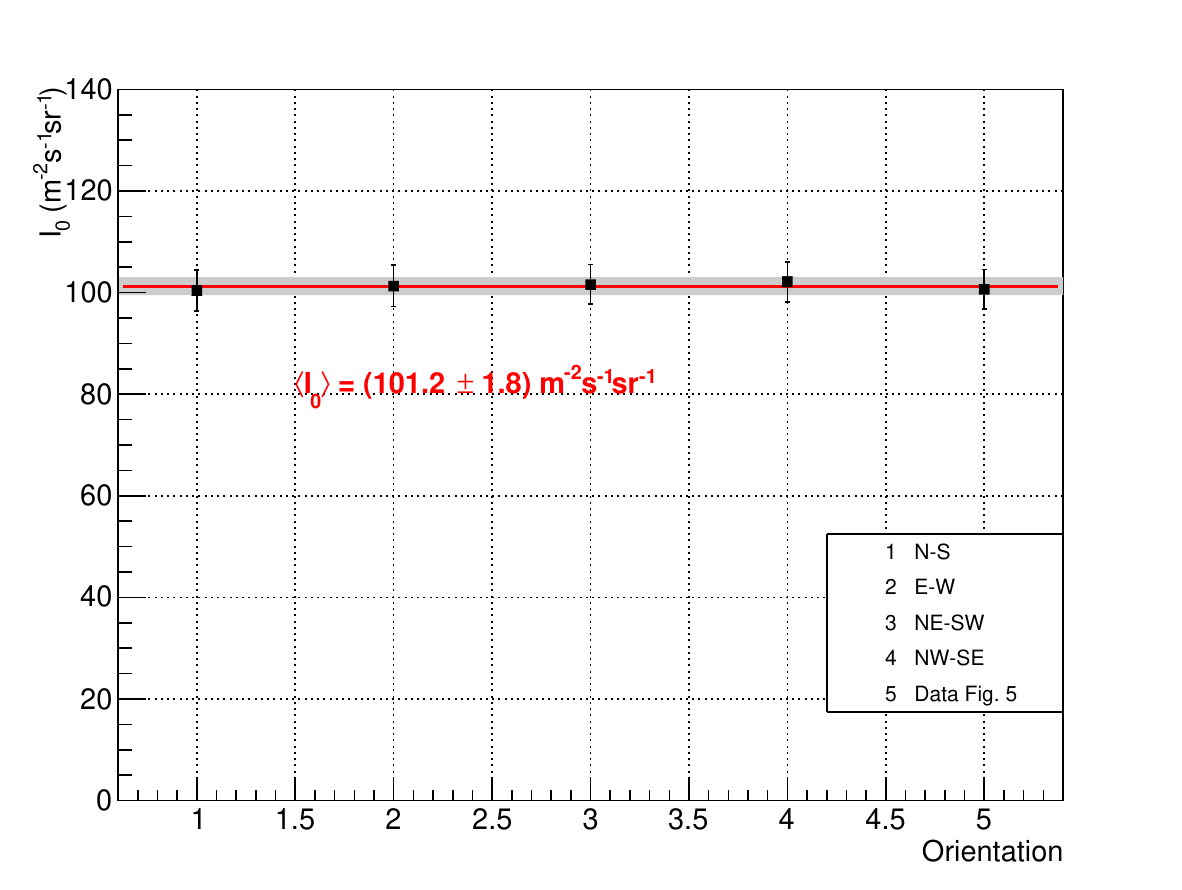}
    \caption{\small Summary of measurements at various azimuthal orientations. The point labeled ``Data Fig. 5'' is oriented 10$^\circ$ wrt. the N-S direction. The red line is a constant fit to the 5 measurements, and the grey band is the error on this fit. In the bottom plot $I_0$ is calculated with the same $\langle n\rangle$ for all the orientations.}
    \label{fig:var-orient}
\end{figure}

\section{\label{sec:var-orient} Varying azimuthal orientations }

The spectrum in Figure \ref{fig:fitted-spectrum} was collected with the scintillator bar aligned at $10^\circ$ with respect to the N-S direction.
Four additional spectra were taken orienting the scintillator bar to lie along the N-S, NE-SW, E-W, and NW-SE directions, each accumulated over 5~hr at approximately the same time of the day and within one week of our initial data spectrum. Weather conditions varied from day to day, from rainy to sunny and temperatures ranging between 20 to 25 $^\circ$C. %
Measurements of $n$ were obtained in the manner described near the end of Section \ref{sec:muon-int}. The top plot in Figure \ref{fig:var-orient} shows the measured values for all 5 orientations, which average $\langle n\rangle =1.90 \pm 0.06$.
The fit was run again over the spectrum for each orientation keeping $n$ fixed to this average value in order to obtain a new measurement of $I_0$ for each case. The uncertainties on the values of $I_0$ for each orientation were obtained by doing the fit with $n$ fixed at the upper and lower limits of the interval for $\langle n\rangle$. The measured values are shown in the bottom plot in Figure~\ref{fig:var-orient}.

It can be seen that the fluctuations in the measurements of $n$ are consistent with the $\sim$10-11\% spread expected from the closure test in Scenario II studied in Section \ref{sec:closure-tests}, and that the uncertainties estimated for each of the fits are also consistent with the means from the 5 measurements and their errors. The precision of the individual measurements does not provide sensitivity to azimuthal anisotropies.

The values of $I_0$ obtained with fixed $n=\langle n\rangle$ are very consistent for all the orientations. Averaging these five measurements we obtain
$I_0=(101.2\pm 1.8({\rm stat})\pm 5.5({\rm syst}))$ $~{\rm m}^{-2}{\rm s}^{-1}{\rm sr}^{-1}$, where we have used the $5.4\%$ systematic error from Table \ref{tab:errors}.
Using the 5\% systematic error found earlier to the measurement of $n$ we report $n=(1.90\pm 0.06({\rm stat})\pm 0.10({\rm syst}))$.

\section{\label{sec:conclusions}Conclusions}

A method to extract the integral vertical intensity and the angular distribution of atmospheric muons   at a given location on the Earth's surface using a stationary plastic scintillator bar detector was presented. 
The method relies on the accurate simulation of the observed deposited energy distribution of atmospheric muons in the detector, for which a muon generator algorithm was developed and coupled with a GEANT4 simulation of the detector response which included the effects of the energy resolution and PMT saturation, as well as a rough description of the laboratory building coverage. The simulated energy spectrum of muons, was fit to an experimentally measured spectrum considering also two exponentially decreasing background components representing the neutrons and electromagnetic showers entering the detector to match the lowest energy region.

Two closure tests were performed to verify the robustness of the fitting procedure and assess the validity of the estimated uncertainties on $I_0$ and $n$. The first test used a simple model without building coverage and no low energy backgrounds. The second test used the full simulation with building coverage and two low energy background components. Within the low statistics of the tests the fits returned the input parameters within the estimated errors. A small remaining systematic offset of order $\sim$4\% towards larger values of $I_0$ and $n$ could not be ruled out but was still expected from the low statistics of the tests. Fitting the data spectrum with the simplified model without building overburden but including the low energy backgrounds produced values of $I_0$ and $n$ about $\sim$8-11\% lower than those obtained including the building coverage.

Using data taken at 5 different azimuthal orientations with 5 hr of exposure in each, and assuming that the angular distribution follows a $\cos^n\theta$ law, we obtain average values of $n=(1.90\pm 0.06({\rm stat})\pm 0.10({\rm syst}))$ and $I_0=(101.2\pm 1.8({\rm stat})\pm 5.5({\rm syst}))$ $~{\rm m}^{-2}{\rm s}^{-1}{\rm sr}^{-1}$ in Mexico City, at the geographical coordinates 19.33$^\circ$N 99.19$^\circ$W, altitude of 2,268 m above sea level, and geomagnetic cut-off of 8.2 ~GV. These values fit reasonably well with expectations for a location at this altitude, and away from strong geomagnetic anomalies. 
Individual fits with a fixed value of $n$, give a small statistical uncertainty on $I_0$ which indicates that this method can be used to track variations in this quantity of the order of or larger than 0.5\% over periods of 5~hr, and is a good option for monitoring applications.

These measurements serve as a proof of principle for the method and show that it could be further developed by combining it with other techniques.

\section*{\label{sec:Ackno}Acknowledgements}
The authors acknowledge the support of DGAPA-UNAM grant number PAPIIT-IN108917, and CONACYT through grants CB-2009/131598 and CB-2014/240666. We also thank ICN-UNAM personnel, Mr. Jos\'e Rangel from the Advanced Manufacturing Laboratory for the machining of a piece to couple the PMTs to the plastic scintillator, and to Ing. Mauricio Mart\'inez Montero for his technical assistance with the instrumentation of the  Detectors Laboratory.



\bibliographystyle{elsarticle-num} 
\bibliography{muVInt-CU-NIMA}



\end{document}